\documentclass[a4paper,fleqn]{cas-dc-arxiv}

\usepackage[numbers]{natbib}
\usepackage{lineno,hyperref}

\usepackage{epsfig}

\usepackage{amssymb}
\usepackage{amsthm}

\usepackage{subfigure}
\usepackage{graphics}
\usepackage{multicol}
\usepackage{graphicx}

\modulolinenumbers[5]

\def\tsc#1{\csdef{#1}{\textsc{\lowercase{#1}}\xspace}}
\tsc{WGM}
\tsc{QE}
\tsc{EP}
\tsc{PMS}
\tsc{BEC}
\tsc{DE}

\begin{document}
\let\WriteBookmarks\relax
\def\floatpagepagefraction{1}
\def\textpagefraction{.001}
\shorttitle{Magnetic properties of clusters of SMPs with central attraction}
\shortauthors{E. V. Novak et~al.}

\title [mode = title]{Magnetic properties of clusters of supracolloidal magnetic polymers with central attraction}



\author[1]{Ekaterina V. Novak}[orcid=0000-0002-4034-6861]
\cormark[1]
\address[1]{Ural Federal University, 51 Lenin av., Ekaterinburg, 620000, Russian Federation.}
\ead{ekaterina.novak@urfu.ru}
\credit{TBD}

\author[1]{Vladimir S. Zverev}[orcid=0000-0003-0422-1436]
\credit{TBD}

\author[1]{Elena S. Pyanzina}
\credit{TBD}

\author[1,2]{Pedro A. S\'anchez}[orcid=0000-0003-0841-6820]
\ead{r.p.sanchez@urfu.ru}
\address[2]{Institute of Ion Beam Physics and Materials Research, Helmholtz-Zentrum Dresden-Rossendorf e.V., D-01314 Dresden, Germany.}
\credit{TBD}

\author[1,3]{Sofia S. Kantorovich}[orcid=0000-0001-5700-7009]
\address[3]{Computational Physics, University of Vienna, Sensengasse 8, Vienna, Austria.}
\ead{sofia.kantorovich@univie.ac.at}

\credit{TBD} 

\cortext[cor1]{Corresponding author}

\begin{abstract}
Supracolloidal magnetic polymers (SMPs) are structures made by crosslinking  magnetic particles. In this work, using Langevin dynamics simulations, we study the zero-field magnetic properties of clusters formed in suspensions of SMPs with different topologies -- chains, rings, X and Y -- that interact via Stockmayer potential. We find that the presence of central attraction, resulting in the formation of large compact clusters, leads to a dramatic decrease of the suspension initial susceptibility, independently from SMP topology. However, the largest decrease corresponds to chain-like SMPs with strongly interacting particles. This is due to the higher rotational degrees of freedom of SMPs with such topology, which allows the particles to reorganise themselves inside the clusters in such a way that their magnetic moments form energetically advantageous vortex structures with negligible net magnetic moments.
\end{abstract}


\maketitle

\sloppy

\section{Introduction}
Since magnetic fluids were synthesised for the first time \cite{resler64a}, the idea of creating magnetic soft smart materials noticeably evolved \cite{2000-zrinyi,2008-baraban,2005-dreyfus,2011-sacanna}. In the early 2000, experimental advances moved the understanding and synthesis of colloidal magnetic particles to a higher level. The same applies to supracolloidal structures based on them. Among the latter, the progress on synthesis techniques of magnetic filaments -- magneto-responsive polymer-like supracolloidal structures made by polymer crosslinking of magnetic particles -- has allowed to decrease their characteristic sizes from several micrometers \cite{2006-poper} to nanometric scale \cite{2015-bharti}. The size of magnetic filaments determines their applications in medicine and technology \cite{2011-devicente,2010-park,2007-matsunaga}.

Magnetic filaments have been actively studied theoretically in recent years \cite{2005-cebers,2009-belovs-pre,2016-wei}. Special attention has been payed to their rheology \cite{2013-chevry,tierno14} and magnetic response \cite{2011-sanchez-sm,2019-kuznetsov}. Recently, using Langevin dynamics computer simulations, we extended the theoretical study of magnetic filaments to systems with more complex crosslinked topologies, known as supracolloidal magnetic polymers (SMPs). The four chosen topologies (chain-, Y-, X- and ring-like backbones) correspond to the most frequently observed self-assembled motifs in systems of magnetic nanoparticles interacting via dipole-dipole interaction \cite{camp00a,2006-klokkenburg,rovigatti11a,2017-ronti}. It is known that, in case of additional central attraction between the dipolar particles (Stockmayer fluids), instead of self-assembling into loose linear or branched clusters, such systems undergo gas-liquid phase transition with particles forming drop-like compact aggregates \cite{1989-smit,1993-vanleeuwen,1992-panagiotopoulos,1995-stevens,1981-adams}.

Here, we want to analyse the presence of a central attraction in systems of chain-, Y-, X- and ring-like SMPs. We expect that crosslinking Stockmayer particles into SMPs will not change the general tendency to phase separate observed in conventional Stockmayer fluids. However, the topology of the SMPs has to affect the structural properties of the formed clusters \cite{2019-nobvak-pre} and, presumably, their magnetic properties.

The manuscript is organised as follows. In the next section we briefly describe models and methods used to investigate SMPs. In the third section we calculate numerically the initial susceptibility for Stockmayer SMPs and explain the observed behaviour by relating it to the distribution of dipoles inside the clusters. Finally, a short summary is provided in Section 4. 

\section{Model and simulation details}\label{sec-model}
We consider SMPs made of monodisperse spherical ferromagnetic particles. In reduced units, such particles have characteristic diameter $\sigma = 1$, mass $m=1$, and carry a point magnetic dipole moment $\vec{\mu}$ at their centers. We account for the long-range magnetic interparticle interactions by means of the conventional dipole-dipole pair potential:
\begin{equation}
U_{dd}(\vec r_{ij})=\frac{\vec{\mu}_{i}\cdot\vec{\mu}_{j}}{r^{3}}-\frac{3\left(\vec{\mu}_{i}\cdot\vec{r}_{ij}\right)\left(\vec{\mu}_{j}\cdot\vec{r}_{ij}\right)}{r^{5}},
\label{eq:dipdip}
\end{equation}
where $\vec \mu_i$ and $\vec \mu_j$ are the dipole moments of particles $i$ and $j$, $\vec r_{ij} = \vec r_i - \vec r_j$ is the displacement vector connecting their centers and $r=\left | \vec r _{ij}\right |$. In this work we sampled two fixed values for the squared dipole moment of each particle in the system, $\mu^2=2$ (weak magnetic interactions) and $\mu^2=5$ (strong magnetic interactions).

For modeling the central attraction between the particles of SMPs we use the Lennard-Jones potential. By taking its energy scale as unity, the latter can be written as follows:
\begin{equation}
U_{\mathrm{{LJ}}}(r)=4 \left ( r^{-12}-r^{-6} \right ). 
\label{eq:LJ}
\end{equation}

\begin{figure}
\centering{\includegraphics[width=0.35\textwidth]{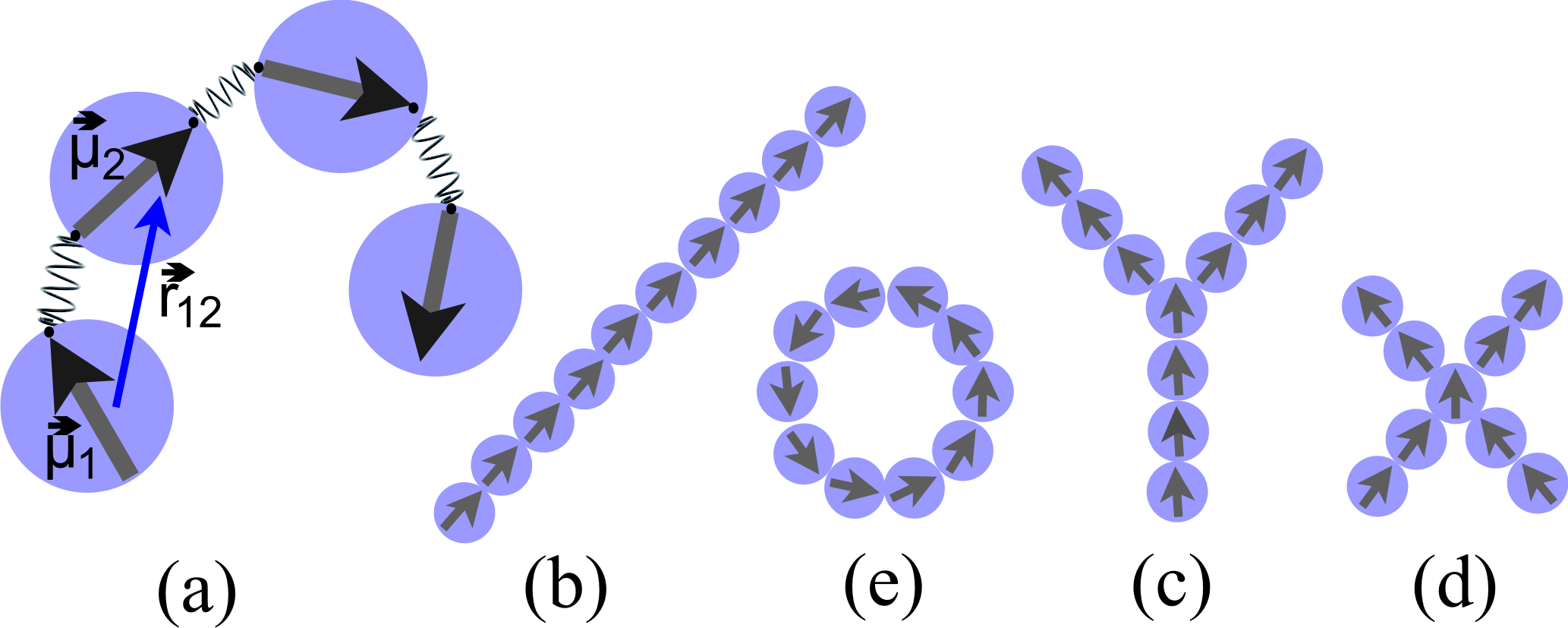}}
\caption{Sketches of the SMP model. (a) Crosslinking and magnetic moments. (b) chain-like SMPs; (c) Y-like SMPs; (d) X-like SMPs; (e) ring-like SMPs}
\label{fig:shapes}
\end{figure}

The bonding between crosslinked particles within every SMP is modeled as a pair potential with two terms. The first term is a simple harmonic spring whose ends are attached to the surface of the bonded particles (see, Fig. \ref{fig:shapes} (a)). The spring attachment points are located at the projection points of the head and the tail of the central dipole moment. The second term corresponds to a FENE potential that limits the maximum extension of the bond. Therefore, the bonding potential is defined as:
 \begin{equation}
     U_{S}(\vec r_{ij}) = \frac{K}{2} \left [ \left( \vec r_{ij} - \frac{1}{2}(\hat{\mu}_i+\hat{\mu}_j) \right )^2 -\frac{r^2_0}{2}\ln\left[1-\left(\frac {\vec r_{ij}}{r_0} \right)^2 \right] \right ],
     \label{eq:bondmodel}
  \end{equation}
where $K$ is the energy scale of the interaction, $\hat{\mu}_i=\vec \mu_i / \left | \vec \mu_i\right |$ and $\hat{\mu}_j=\vec \mu_j / \left | \vec \mu_j\right |$ are the unitary vectors parallel to each associated dipole moment and $r_0$ is the maximum allowed extension for the bond. According to our previous studies\cite{2017-novak-jmmmb, rozhkov17a}, we take $K=30$ and $r_0=1.5$ in our reduced units.

We performed molecular dynamics simulations in the canonical ensemble at reduced temperature $T=1$, using a Langevin thermostat in order to implicitly take into account the effects of thermal fluctuations of the carrier fluid. The systems consisted of $N_{SMP} = 512$ identical SMPs with size either $L = 10$ for chain-, Y- and ring-like SMPs or $L=9$ for X-structures. Sketches of the bonding model and SMP topologies are shown in Figs.~\ref{fig:shapes} (b)-(e). We used a cubic simulation box of volume $V$ with periodic boundary conditions. The number density $\rho = N_{SMP} L/V$ was fixed to 0.05. 

The simulations were performed with the ESPResSo 3.2.0 package \cite{2006-limbach}. The system was first equilibrated at high $T=4$ to assure random distribution of SMPs before switching on magnetic interactions and central attraction. Afterwards, before the production runs were performed, the system was re-equilibrated at $T=1$ for $9 \cdot 10^5$ integration steps, using a time step $\delta t =5 \cdot 10 ^{-3}$. Finally, a production cycle of $3 \cdot 10^6$ steps was performed, in which the system configurations were recorded at intervals of $10 ^5$ steps. Magnetic interactions were calculated using the dipolar-P$^3$M algorithm \cite{2008-cerda-jcp}. Further details of the simulation protocol can be found in reference \cite{rozhkov17a}.

\section{Results}\label{sec-res}

\begin{figure}
\centering{\includegraphics[width=0.38 \textwidth]{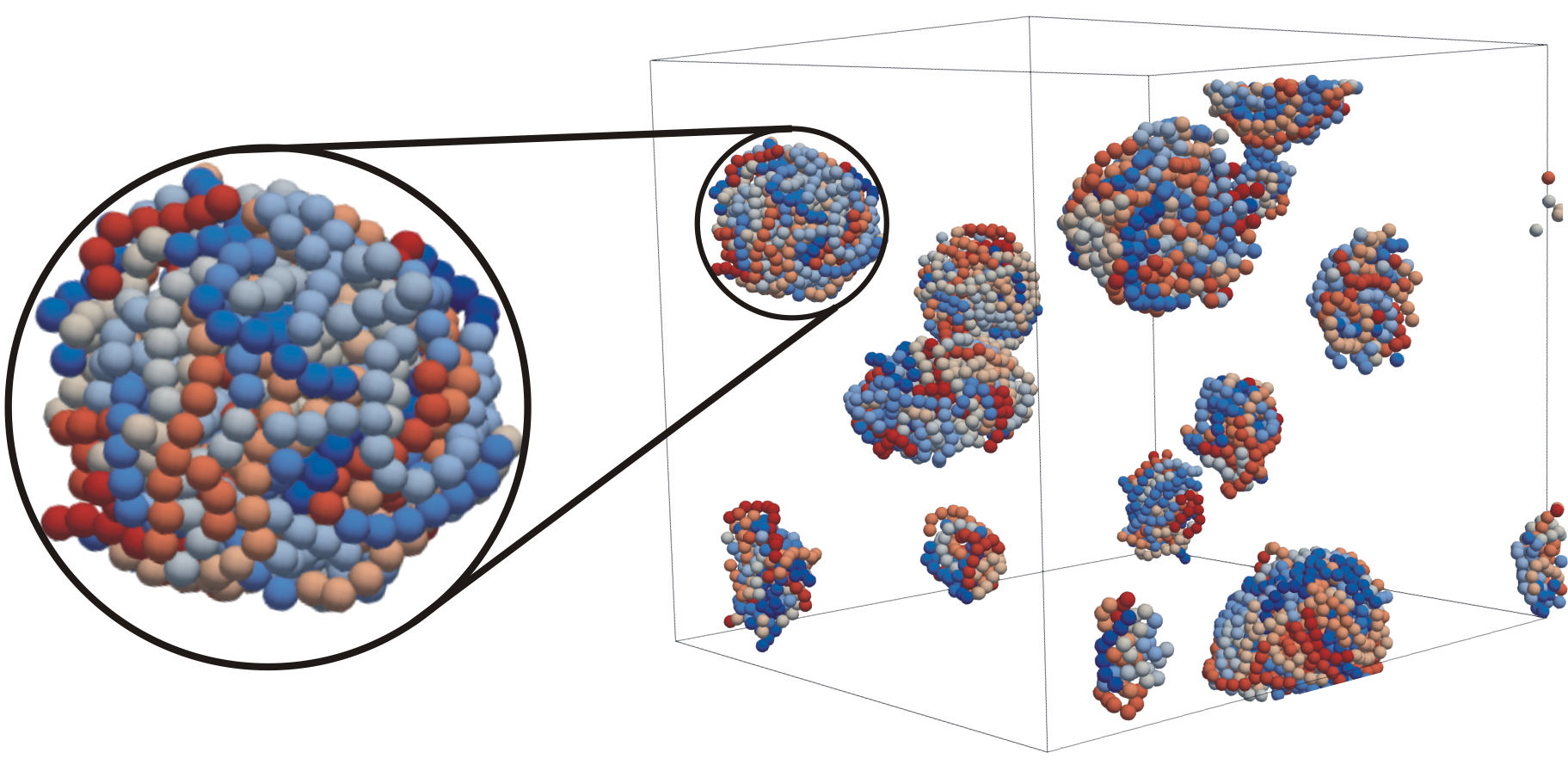}}
\caption{Typical snapshot obtained for a system of 512 X-like SMPs with $\mu^2=5$. Each SMP has a difefrent colour. One of the clusters is enlarged for clarity.}
\label{fig:snapshot}
\end{figure}

A typical snapshot with zoomed-in cluster is presented in Fig. \ref{fig:snapshot}. One can see that the clusters are quasi-spherical and rather compact. 
Such a compact structure of the clusters, led by the central attraction, in combination with the presence of permanent crosslinks that couple particle rotations the SMP backbone, suggests that magnetising these clusters might be more difficult than with individual SMPs or particles. In order to check this conjecture, we compute numerically the initial static magnetic susceptibility on the basis of the fluctuation--dissipation theorem \cite{1951-callen}: 
\begin{equation}
\chi=\frac{4\pi\rho}{3T}
\left(
\langle  \left| \boldsymbol{M} \right|^2 \rangle - 
\left| \langle \boldsymbol{M} \rangle \right|^2
\right),
\label{eq:ISMS}
\end{equation}
where the relative net magnetic moment of each SMP, $M$, is determined according to:
\begin{equation}
M=\frac{1}{L\mu}\left|\sum_{i=1}^L\vec{\mu}_i\right|.
\label{eq:magmom}
\end{equation}
The results are summarised in Table \ref{table:in-susc}. Here we collected the values of initial susceptibility not only for suspensions of Stockmayer SMPs (S-SMPs), but also for suspensions of SMPs without central attraction (SMPs) and for non-crosslinked conventional systems of dipolar soft spheres (DSS). All numbers in this table are normalised to Langevin susceptibility $\chi_L = 4\pi\rho\mu^2/3$, that corresponds to an ideal superparamagnetic gas. Additionally, in the fourth and the seventh columns we provide the values of the ratio between the susceptibility of Stockmayer SMPs and that of SMPs without central attraction., $\kappa = \chi_{S-SMP}/\chi_{SMP}$.
\begin{table*}
\centering
\small
\begin{tabular}{c||ccc|ccc}   \hline\hline
     &  & $\mu^2 = 2$  &   &  & $\mu^2 = 5$ &  \\\hline
 & S-SMPs  & SMPs & $\kappa$ & S-SMPs  & SMPs & $\kappa$ \\ \hline
C & 0.34 & 3.59 & 10.56 & 0.11 & 6.70 & 60.91\\
Y & 0.39 & 2.86 & 7.33 & 0.14 & 5.02 & 35.86\\
X & 0.42 & 1.80 & 4.29 & 0.14 & 3.25 & 23.21\\ 
R & 0.10 & 0.10 & 1.00 & 0.06 & 0.04 & 0.66\\ \hline
DSS &  & 1.17& &  & 3.34  &\\ \hline
\end{tabular}
  \caption{Initial magnetic susceptibility divided by Langevin susceptibility calculated for the following systems.  C -- chain-like SMPs,  Y -- Y-like SMPs, X -- X-like SMPs, R -- ring-like SMPs, and non-crosslinked systems of dipolar soft spheres DSS. Prefix ``S'' denotes Stockmayer. $\kappa$ shows the ratio between values for susceptibilities of S-SMPs to those of SMPs (without central attraction).}
  \label{table:in-susc}
\end{table*}
One can clearly see that the initial susceptibility of compact clusters of S-SMPs is one order of magnitude smaller than that of SMPs without central attraction for all but ring-like topologies. In case of rings, the susceptibility is very low as the magnetic flux closure takes place inside the individual SMPs, analogously to the effect observed in \cite{2013-kantorovich-prl}.
\begin{figure*}
	\centering
\subfigure[]{\label{fig:vort-C}\includegraphics[width=0.32 \textwidth]{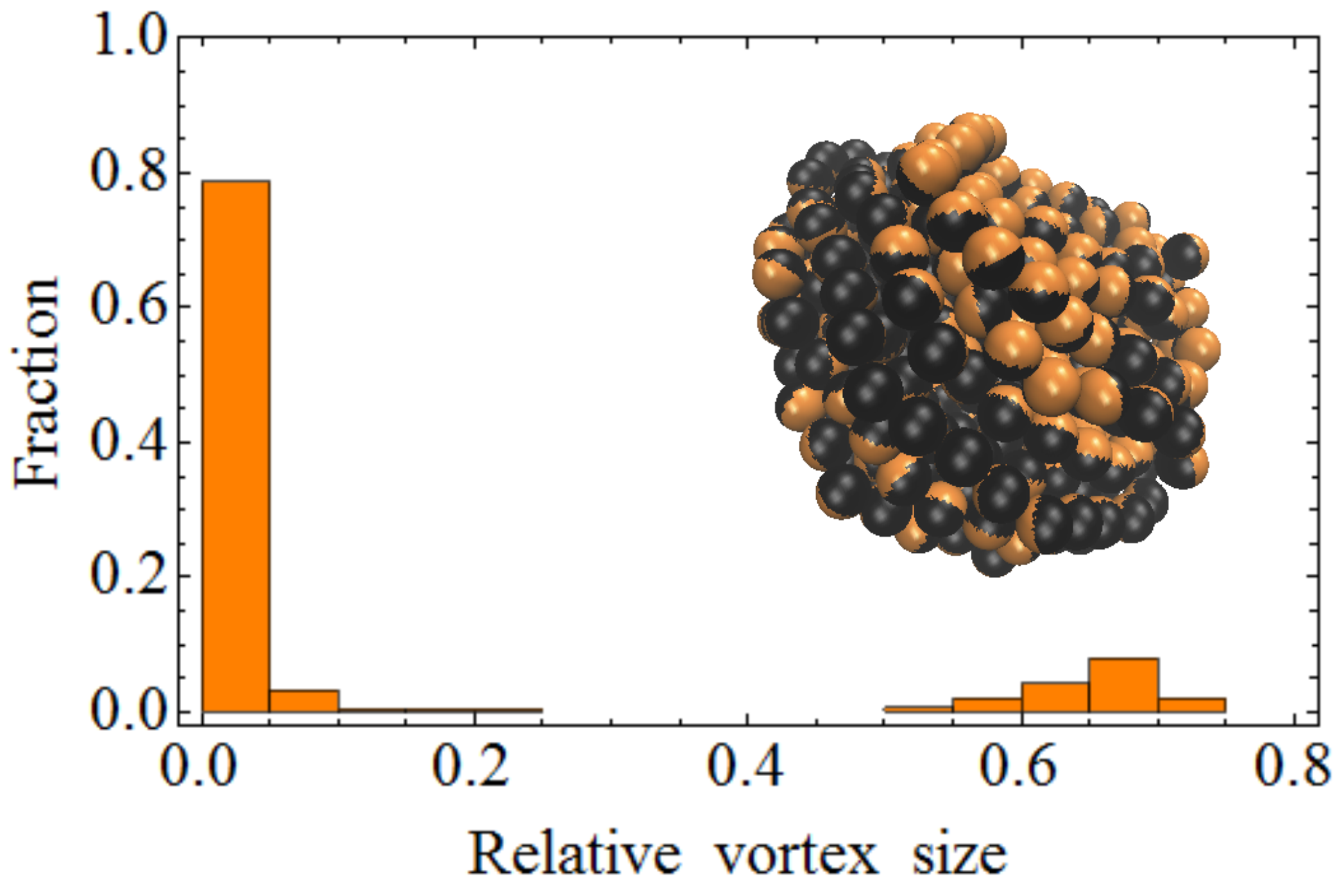}}\hspace{0.3cm}
\subfigure[]{\label{fig:vort-Y}\includegraphics[width=0.32 \textwidth]{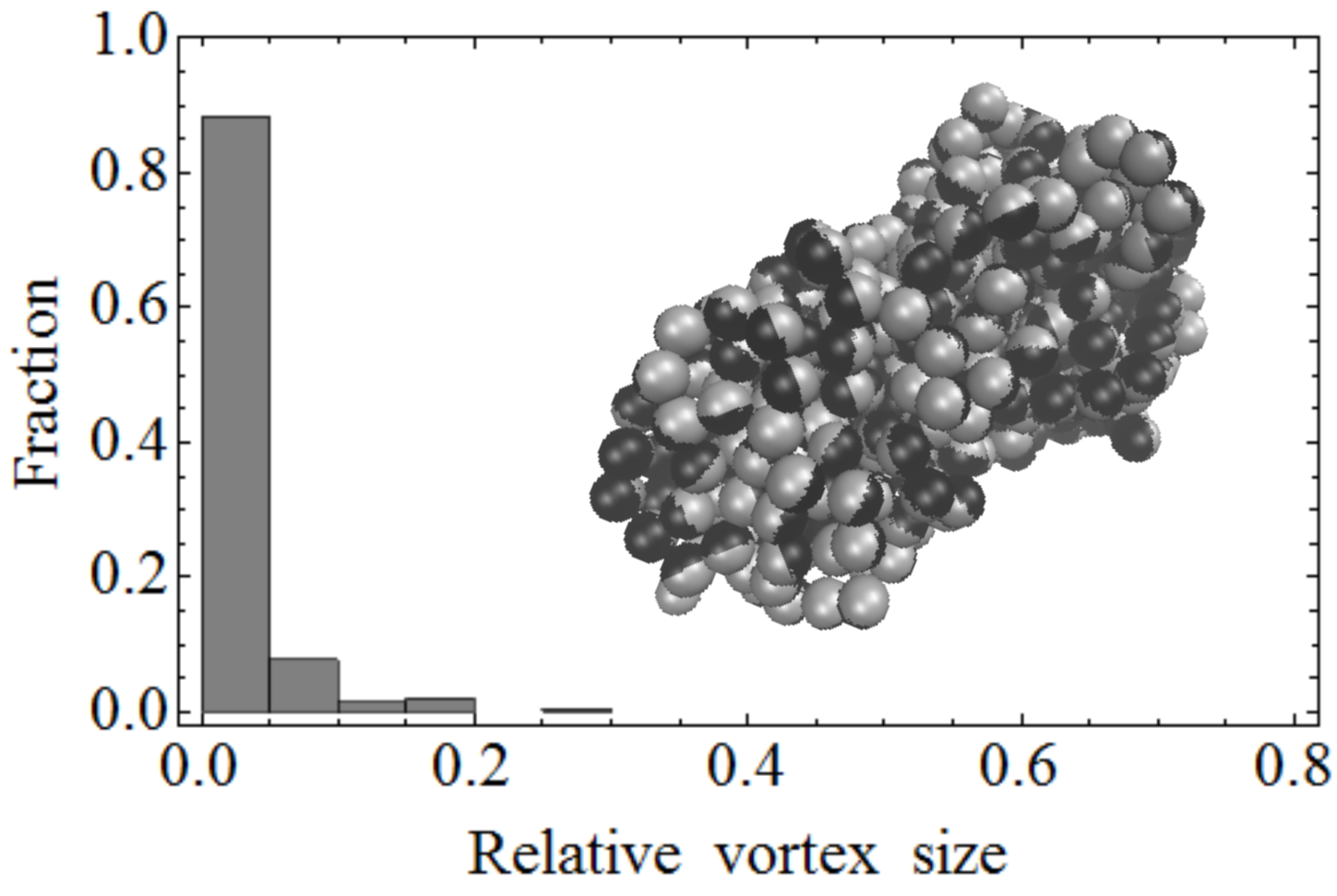}}\\
\subfigure[]{\label{fig:vort-X}\includegraphics[width=0.32 \textwidth]{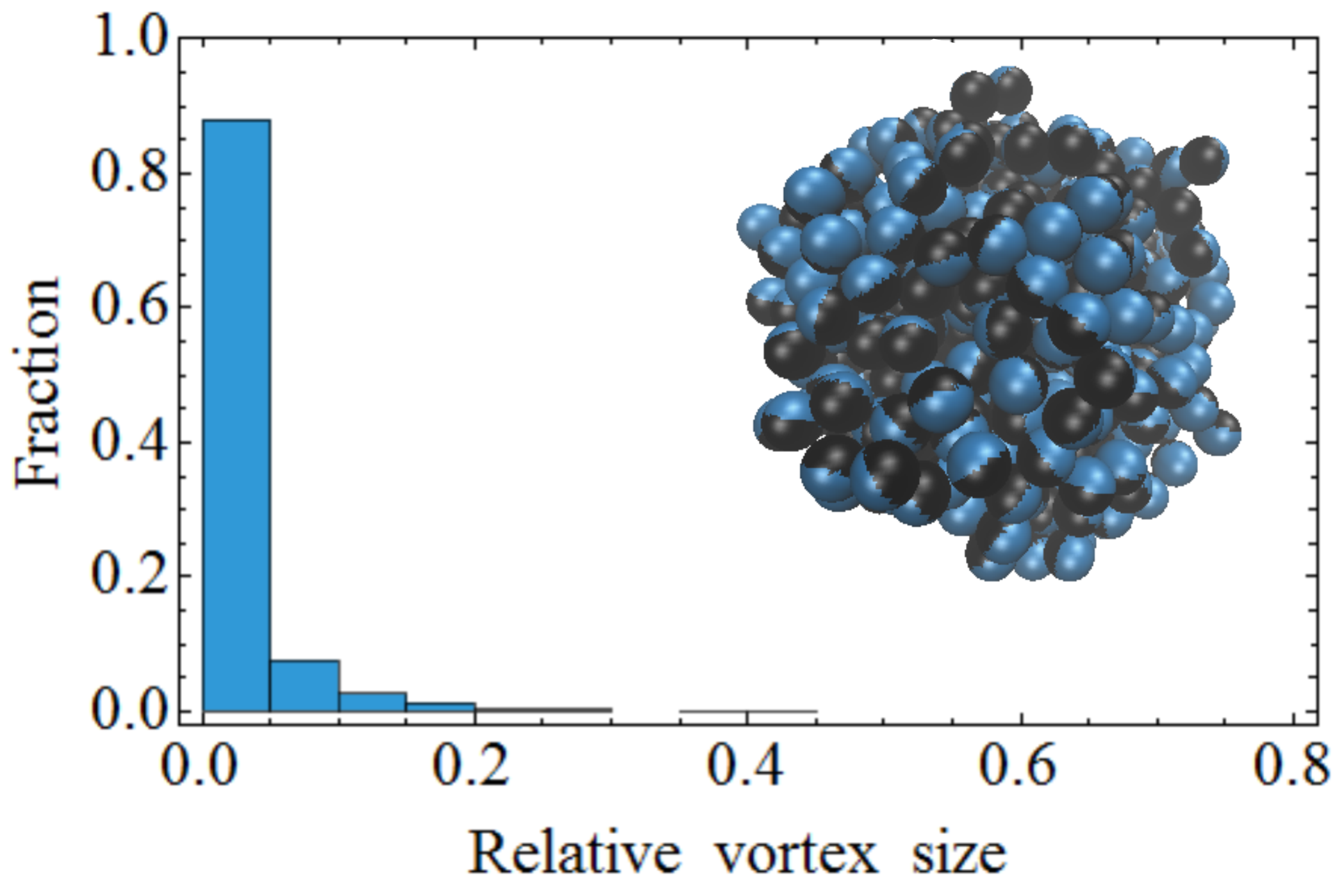}}\hspace{0.3cm}
\subfigure[]{\label{fig:vort-R}\includegraphics[width=0.32 \textwidth]{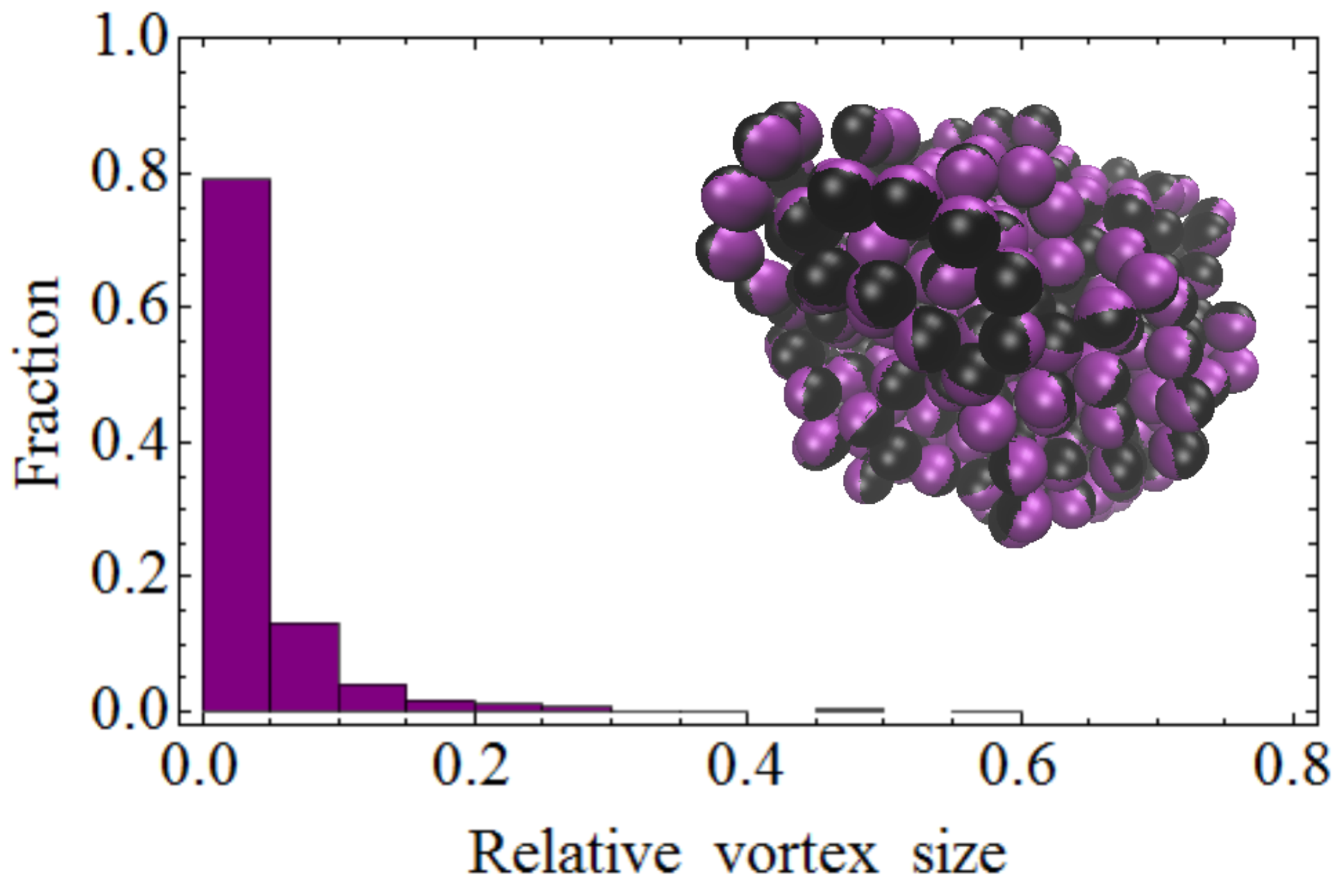}}
	\caption{Histograms of relative vortex size and typical snapshots showing the direction of the curl at the position of each particle, indicated by two-coloured light and dark hemispheres, corresponding to $\mu^2=5$. (a) For chain-like SMPs; (b) Y-like SMPs; (c) X-like SMPs; (d) ring-like SMPs.}
	\label{fig:vort-hist}
\end{figure*}
Moreover, there are three important tendencies in the behaviour of chain-, Y- and X-like SMPs to be noticed in Table \ref{table:in-susc}. First, comparing the values of $\kappa$ for different $\mu^2$, one can see that the decrease of the susceptibility due to the formation of compact clusters is much larger for strongly interacting particles.  Besides that, for $\mu^2=2$, any crosslinking leads to an increase in susceptibility. This is related to the fact that in the system of noncrosslinked dipolar soft spheres with $\rho=0.05$ and $\mu^2 = 2$, the particles are rather weakly correlated, as it can be seen from the value of the initial susceptibility which is basically equal to that of Langevin. If, instead, $\mu^2=5$, crosslinking in X-like SMPs leads to a slight decrease in the initial susceptibility: each linear segment of an X-like SMP made of 9 magnetic particles has only two beads. The average chain length in the system of DSS, see for example \cite{wang_pre}, is around 2.5. Thus, in this case, the crosslinking weakens the correlations. Clearly, the strongest enhancement of the interparticle correlations is observed for chain-like topology. At the same time, for the latter, the value of $\kappa$ turns out to be the largest.

In order to explain this, we thoroughly analysed the distribution of dipoles inside the clusters. Such distribution defines a discrete vector field, $F$ (not to be confused with the magnetic field created by the distribution of dipoles). $F$ is given by the position of the centre of each particle and the vector of its magnetic moment. By treating the distribution of dipole moments as a vector field one can calculate its curl. The latter can show if there are vortex-like arrangements of dipoles and if the number and size distribution of these vortices depends on the constituent SMP topology. This is important because vortex-like structures, possesing a low net magnetic moment, can be responsible for drastic decrease of the susceptibility. We calculated curls for each cluster found in simulations on a grid provided by particle centres using the weighted regression method due to its robustness \cite{2003-petrovskaya,2011-correa}. The calculations were performed for all clusters whose size was in a range between 45 and 48 SMPs. As 3D vector plots are very difficult to visualise, we processed the curl vectors further: for each curl vector we found the neighbourhood (neighbouring particles) in which the orientation of the curls of two adjacent particles deviate by less than 15\% ($\sim \pi/6$). If the neighbourhood contained less than four particles it was discarded, otherwise it was named as a vortex of a given size. Additionally, the vortex size was normalised by the total number of particles in the cluster. With this procedure, we calculated the histograms of relative vortex sizes for solutions of different SMPs. The results for  $\mu^2=5$ are presented in Fig.~\ref{fig:vort-hist}. It is clearly seen that for chain-like SMP clusters, the size distribution of vortices is bimodal (Fig. \ref{fig:vort-C}). For each distribution, we also include a typical cluster with particles showing two different light/dark coloured hemispheres, so that the division plane is perpendicular to the curl in this point. Visual inspection of the cluster snapshot in the inset of Fig. \ref{fig:vort-C} reveals the presence of a large neighbourhood of particles turned with their dark side to the reader. The curl vectors calculated in the centres of these particles are almost coaligned, thus, these particles are forming a large vortex. In none of the other three Figs. \ref{fig:vort-Y} -- \ref{fig:vort-R} one finds such large vortices, however, there are many smaller local ones. For clusters made of ring-like SMPs, some of the vortices coincide with the rings themselves, as it can be seen in the inset of Fig. \ref{fig:vort-R}. The absence of large vortices in the clusters of Y-, X- and ring-like SMPs has a simple geometrical explanation: in Fig. \ref{fig:shapes} (c) -- (e) the dipoles are shown in the crosslinked orientation and any deviation from this orientation is penalised by elastic forces, thus making their rotational degrees of freedom on average more constrained than for chain-like SMPs (Fig. \ref{fig:shapes} (b)).   

\section{Conclusions}
In the present study we analysed zero-field magnetic properties of suspensions composed of Stockmayer SMPs of four different topologies. We found that, due to the formation of compact clusters, the overall initial magnetic susceptibility decreases dramatically for chain-, Y- and X-like topologies. However, this decrease for chain-like SMPs is much stronger than that of other topologies, especially for systems with strong magnetic interactions. This can be explained by calculating the curls of magnetic moment distributions. It turned out that chain-like SMPs, when self-assembling in large compact clusters, rearrange their particle dipole moments inside the clusters to form large vortexes, whose total magnetic moment is negligible. Other SMPs cannot exhibit the same behaviour due to the constraints on the particle rotations away from the crosslinked backbone. This result might be of importance for the behaviour of Stockmayer SMP clusters in flow under the applied magnetic fields, particularly if one aims, for instance, at optimising transport properties.

\section{Acknowledgements}
Research supported by the Russian Science Foundation Grant No.19-72-10033. S.S.K. acknowledges support from the Austrian Research Fund (FWF), START-Projekt Y 627-N27.

\end{document}